\documentclass[11pt,twoside]{article}
\usepackage{CAGN2021}
\usepackage{graphicx}
\usepackage{amsmath}
\usepackage{breqn}

\usepackage[T1]{fontenc} % Computer Modern (CM) fonts

\usepackage{latexsym}
\usepackage{verbatim}

\usepackage{ifpdf}  
\ifpdf  
      \DeclareGraphicsExtensions{.pdf,.png,.jpg}  
\else  
      \DeclareGraphicsExtensions{.eps}  
\fi 

\newcommand{\fado}{\textsc{Fado}}\newcommand{\starl}{\textsc{Starlight}}
\newcommand{\stec}{\textsc{Steckmap}}\newcommand{\reb}{\textsc{Rebetiko}}
\newcommand{\Con}{{\tt CONT}}\newcommand{\Tau}{{\tt TAU1}}
\newcommand{\logtL}{$\left<log(t_\ast)\right>_{L}$}\newcommand{\logtM}{$\left<log(t_\ast)\right>_{M}$}
\newcommand{\logZL}{$\left<log(Z_\ast)\right>_{L}$}\newcommand{\logZM}{$\left<log(Z_\ast)\right>_{M}$}

\begin{document}
% please do not un-comment the next line
% \input{../../proceeding-book/expages.tex}\setpagenumber{1}

\vskip 1.0cm
\markboth{C.~Pappalardo et al.}{Spectral synthesis with FADO}
\pagestyle{myheadings}
%
%%%%  USE THE LINE THAT DESCRIBES THE CHARACTER OF YOUR WORK %%%%%%
%
\vspace*{0.5cm}
\parindent 0pt{Contributed  Paper}
%\parindent 0pt{Poster}

%\vskip 0.3cm

\vspace*{0.5cm}
\title{Self-consistent population spectral synthesis with FADO: mean stellar metallicity of galaxies in spectral synthesis methods}

\author{Ciro Pappalardo$^{1,2}$}

\affil{$^1$Instituto de Astrof\'{i}sica e Ci\^{e}ncias do Espa\c{c}o, Universidade de Lisboa - OAL, Tapada da Ajuda, PT1349-018 Lisboa, Portugal\\
$^2$Departamento de F\'{i}sica, Faculdade de Ci\^{e}ncias da Universidade de Lisboa, Edif\'{i}cio C8, Campo Grande, PT1749-016 Lisboa, Portugal}

\begin{abstract}
In this work, we investigate the reliability of spectral synthesis methods in the estimation of the mean stellar age and metallicity, addressing the question of which signal-to-noise ratios (S/N) are needed to determine these quantities and how these depend on the tool used to model the spectra. To address this problem we used realistic simulated spectra containing stellar and nebular emission, reproducing the evolution of a galaxy for a constant and exponentially declining star formation law. The spectra have been degraded to different S/N and analysed with three different spectral synthesis codes: \fado, \stec, and \starl\ assuming similar fitting set-ups and the same spectral bases.

For S/N $\le5$ all tools considered show a large diversity in the results. \fado\ and \starl\ find median differences in light-weighted mean stellar ages of $\sim$0.1 dex, while \stec\ shows a higher value of $\sim$0.2 dex. For S/N $> 50$ the median differences in \fado\ are $\sim$0.03 dex ($\sim7$\%), a factor 3 and 4 lower than the 0.08 dex ($\sim$20\%) and 0.11 dex ($\sim$30\%) obtained from \starl\ and \stec, respectively. 

Our results indicate that phases of high specific star formation rate (sSFR) in galaxies require analysis tools that do not neglect the nebular continuum emission in the fitting process, since purely stellar models would have strong problems in the estimation of star formation history, even in presence of high S/N spectra. The median value of these differences are of the order of 7\% (\fado), 20\% (\starl), and 30\% (\stec) for light-weighted quantities, and 20\% (\fado), 60\% (\starl), and 20\% (\stec) for mass-weighted quantities. That implies a severe overestimation of the mass-to-light ratio and stellar mass, even in the presence of a mild contribution from the nebular continuum. Our work underlines once more the importance of a self-consistent treatment of nebular emission, which is the only route towards a reliable determination of the assembly of any high-sSFR galaxy at high and low redshift.

 
\end{abstract}

\section{Introduction}

Galaxy evolution will undergo a leap forwards in the next decade when a new generation of multi-object spectroscopic facilities will be operational. Surveys like 4MOST \citep{dej}, WEAVE \citep{dal}, and MOONS \citep{cir,cir2,mai} will open a new window into an unexplored area of research. These spectra will tackle key problems related to the galaxy assembly, shedding light on the physical processes governing their evolution. It is then of paramount importance to refine and test reliable tools able to infer the properties of a galaxy from the analysis of medium- or high-resolution spectra.
To address these questions, in this work we investigate the limits and reliability of different spectral synthesis methods in the estimation of two specific quantities representatives of the evolutionary status of a galaxy: the mean stellar age and mean stellar metallicity. These parameters are fundamental for determining the assembly history of a galaxy, by providing key insights into its star formation history (SFH) and stellar mass growth.

The main question we want to address is at which signal-to-noise (S/N) values it is still possible to determine, from a galaxy spectrum, its mean stellar age and metallicity and how this depends on the tool used to model the spectra.

To do this we built a set of simulated spectra reproducing the evolution of a galaxy for a continuous star formation and a single initial burst with an exponentially declining star formation law. We then degraded the synthetic spectra to different S/N, simulating different observing conditions. The mock spectra produced were then analysed with three spectral synthesis codes: \fado, \stec, and \starl\ assuming similar initial set-ups and spectral bases. The main goals are a) to determine the S/N where the results for the parameters considered are still reliable; b) to identify which stellar ages the fitting tools are more sensitive to, and c) to quantify the effect of nebular emission component in population synthesis codes. Section~\ref{procedure} describes the procedure used, while Section~\ref{results} shows our results. In Section~\ref{discussion} we present our conclusions.

\section{Procedure}
\label{procedure}

\begin{figure*}     \centering
     \includegraphics[clip=,width=0.49\textwidth]{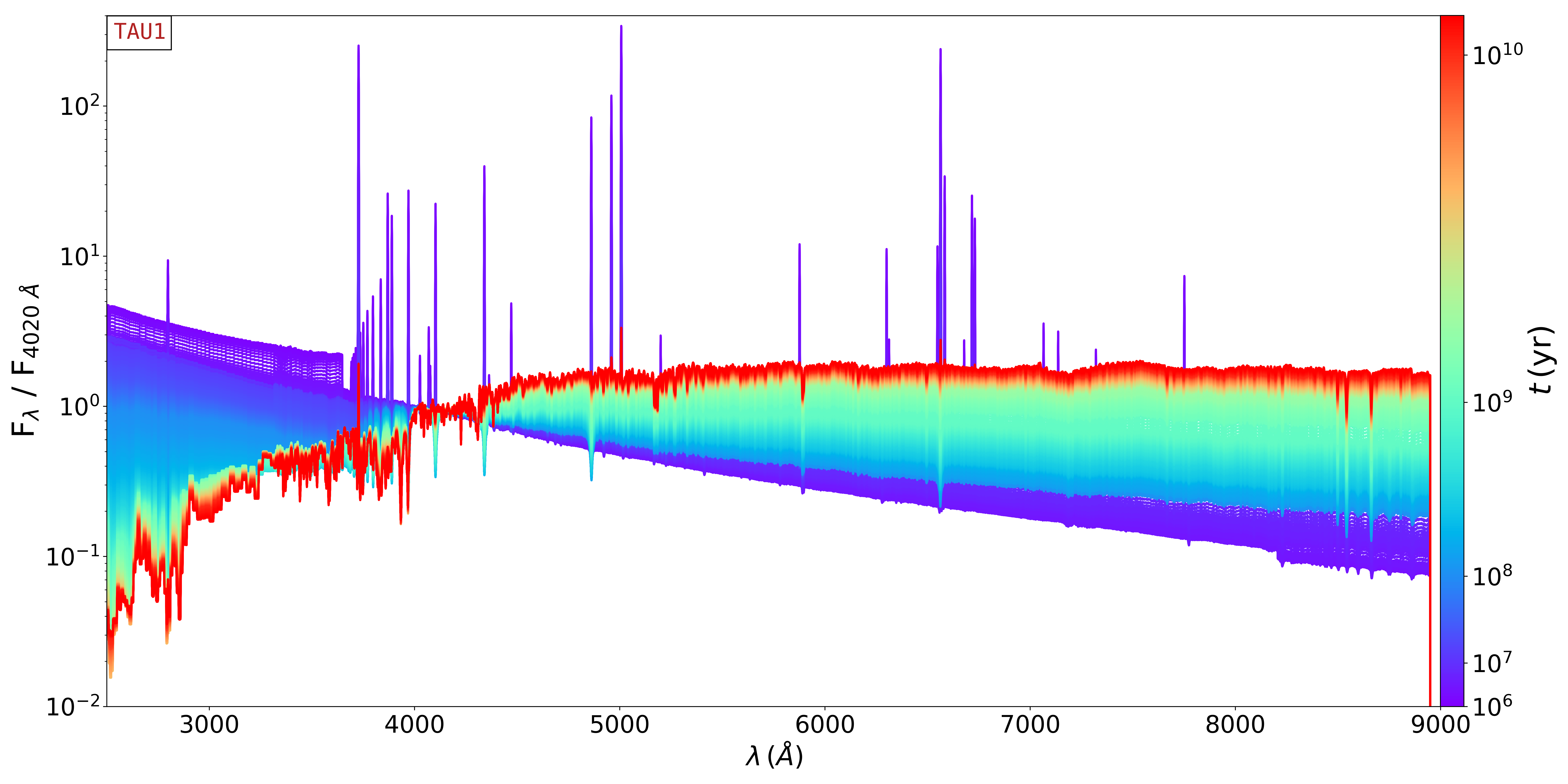}
     \includegraphics[clip=,width=0.49\textwidth]{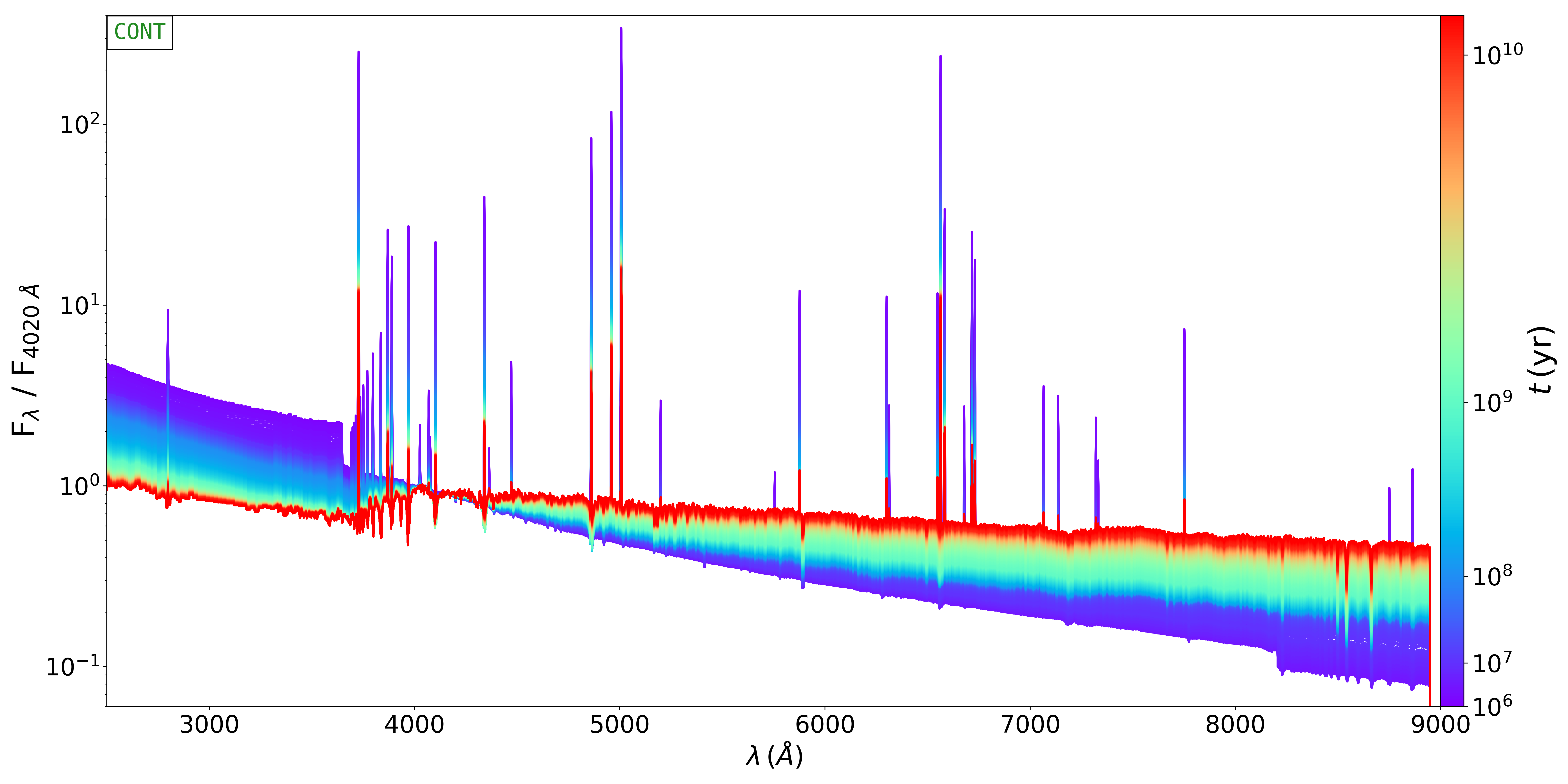}     
     \caption{Synthetic spectra normalised at $\lambda$ = 4020 \AA\ for \Tau\ (instantaneous burst, left panel) and \Con\ (continuous, right panel) SFH models. Figures adapted from Fig. 1 and Fig. 2 of \citealt{car}.}
     \label{rebetikospectra}
 \end{figure*}

 The {\it Reckoning galaxy Emission By means of Evolutionary Tasks with Input Key Observables} (\reb) code built the synthetic spectra analysed in this work, applying the following equation:
 
 \begin{equation}\begin{split}
     F(\lambda,t)= \int_0^t \Phi(t-t') & F^{SSP}_{\lambda}(t',Z)dt'
     \\
     & +\frac{\gamma_{eff}(T_e)}{\alpha_B(T_e)}\int_0^t \Phi(t-t')q^{SSP}(t',Z)dt'
     \label{eq1}
 \end{split}.\end{equation}
 \noindent
Eq. \ref{eq1} reproduces the evolution of a galaxy spectral energy distribution (SED) as a function of time. For each $t$ the flux $F(\lambda,t)$ at a specific wavelength $\lambda$ depends on two terms: The first term represents the emission due to the stellar component, which is the integral of the different SSPs weighted for the star formation rate (SFR; $\Phi$). The second term represents the flux contribution due to the nebular continuum, which depends on the number of hydrogen ionising photons $q^{SSP}$ weighted for the star formation. The spectra assume a constant solar metallicity and two different star formation histories (SFHs): a continuous star formation ({\tt Cont} model), and an instantaneous burst followed by an exponentially declining star formation law with $\tau$ = 1 Myr (\Tau\ model). The \reb\ spectra for the \Con\ and \Tau\ models are shown in Fig. \ref{rebetikospectra} (adapted from \citealt{car}).
 
 %---
 %-----

The evolution of the synthetic SEDs simulated in \reb\ is characterised by the mean stellar age and the mean stellar metallicity weighted by the light (mass) contribution to the total galaxy light (mass), considering logarithmic time-steps, as in \citealt{cid}:
 \begin{equation}\begin{split}
   \left<log(t_\ast)\right>_L = \sum_{j=1}^{N_\ast} x_j \cdot log (t_j),
  \\
   \left<log(t_\ast)\right>_M = \sum_{j=1}^{N_\ast} \mu_j \cdot log (t_j)
 \end{split}\label{sspAge}.\end{equation}
 
 Analogously for the metallicity,
 
 \begin{equation}\begin{split}
   \left<log(Z_\ast)\right>_L = \sum_{j=1}^{N_\ast} x_j \cdot log (Z_j),
  \\
   \left<log(Z_\ast)\right>_M = \sum_{j=1}^{N_\ast} \mu_j \cdot log (Z_j).
 \end{split}\label{sspZ2}\end{equation} 
 
 \noindent
 In Eq. \ref{sspAge} and \ref{sspZ2}, $t_j$ ($Z_j$) represents the age (metallicity) of the j-th SSP element, where $x_j$ and $\mu_j$ are their associated light and mass fractions at 4020 \AA, respectively.
 
 We add to the input \reb\ spectra random Gaussian noise with a varying standard deviation, obtaining a set of spectra with constant S/N values of 3, 5, 10, 20, 50, and 100. The spectral basis for the analysis is built from the {\tt BaseL} of \citealt{bru}, considering 100 SSPs, 25 ages and four metallicities ($Z$ = 0.004, 0.008, 0.02 and 0.05).

 We selected for the analysis three spectral synthesis tools: \fado\ \citep{gom}, which includes in its analysis nebular emission; the purely stellar code \starl\ \citep{cid}; and \stec\ \citep{ocv}, which can be considered a `hybrid' approach.

\section{Results}
\label{results}

The synthetic spectra built with \reb\ were analysed with the tools described in Section \ref{procedure}, for six S/N values in the range $3 \le$ S/N $\le$ 100. For each SFH, we estimated the mean stellar age and metallicity, weighted by mass or light, as defined in Eq. \ref{sspAge} and \ref{sspZ2}. 
  
  \subsection{Continuous star formation: \Con\ models}
  
\begin{figure*}     \centering
     \includegraphics[clip=,width=0.99\textwidth]{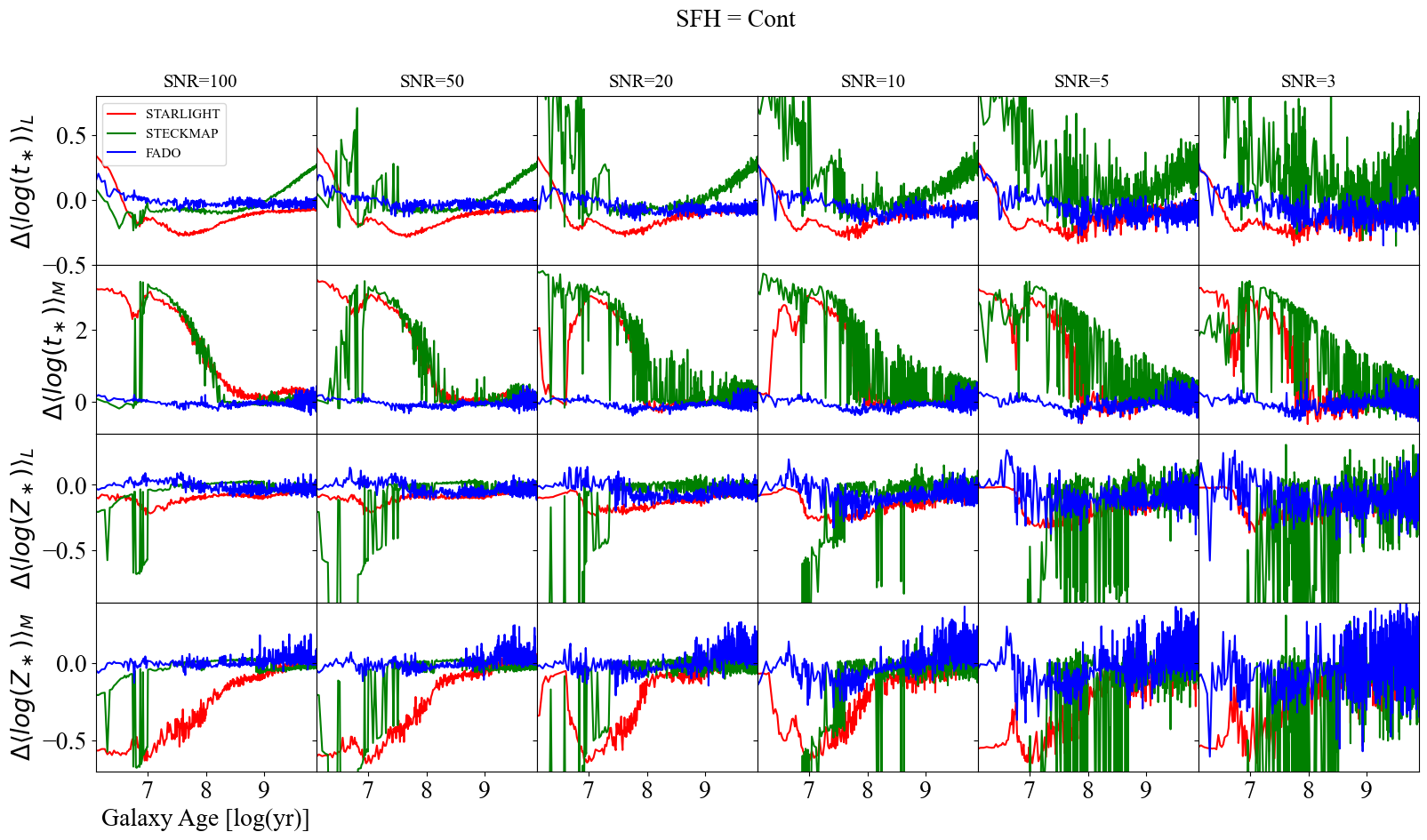}     
     \caption{Results of \Con\ model analysis. Luminosity- and mass-weighted mean stellar age (first and second row), and metallicity (third and fourth row) residuals for \fado\ (blue line), \starl\ (red line), \stec\ (green line) as a function of the galaxy age. Luminosity-weighted quantities for \fado\ and \starl\ are obtained considering the best-fitting populations vector at 4020 \AA. On top of each panel the S/N of the input \reb\ spectrum are reported. {\it Credits: Reproduced with permission from Astronomy \& Astrophysics, \textsuperscript{\textcopyright} ESO.}}
     \label{Cont-logtL}
 \end{figure*}

  Fig. \ref{Cont-logtL} shows the difference between \logtL, \logtM, \logZL, and \logZM\ for \Con\ models with respect to the input \reb\ value. All tools show large differences of the fit at S/N = 3 and have problems at ages $\log(t/\mathrm{yr}) < 8$, where the results have a large spread as a consequence of the instability of the solutions. For older ages, the results are more stable, with a tendency to underestimate \logtL.
  
  For S/N $> 20$ the differences between the \reb\ \logtL\ and the best fits from \fado\ are $\sim$0.03 dex ($\sim7$\%), a factor 3 and 4 lower than the 0.08 dex ($\sim$20\%), and 0.12 dex ($\sim$30\%) obtained from \starl\ and \stec. The fits show small variations, and the $\Delta$\logtL\ at ages above $\log(t/\mathrm{yr}) \sim 8$ decreases gradually, in correspondence with the predominance of the post-AGB stars.
  
  Once input spectra have S/N above 10 and galaxy models have ages above $\log(t/\mathrm{yr}) > 8$ all tools converge towards the input solar metallicity. \starl\ shows some discrepancies, $\Delta$\logZL$\sim 0.2$ dex, for ages below $\log(t/\mathrm{yr}) < 8$, with similar patterns at S/N = 50 and 100. \stec\ shows a mild overestimation of the metallicity at higher S/N and old ages, lower than 0.01 dex.

  Mass-weighted quantities are more noisy with respect to light-weighted fits (Fig. \ref{Cont-logtL}), with \fado\ showing at S/N $\ge 10$ uncertainties of $\sim$0.09 dex, larger than the 0.04 dex obtained for \logtL. \starl\ results show median discrepancies of 0.12-0.24 dex with no real trend with increasing S/N. All the tools have problems with the metallicity evolution for spectra at S/N $<$ 20: \fado\ shows more stable fits at higher S/N, while \starl\ underestimates the metallicity by $\sim$0.01 dex, above all at young ages.
 
   %------------------------
 %--------------------------- 

 %------------------------
 \begin{figure*}     \centering
     \includegraphics[clip=,width=0.99\textwidth]{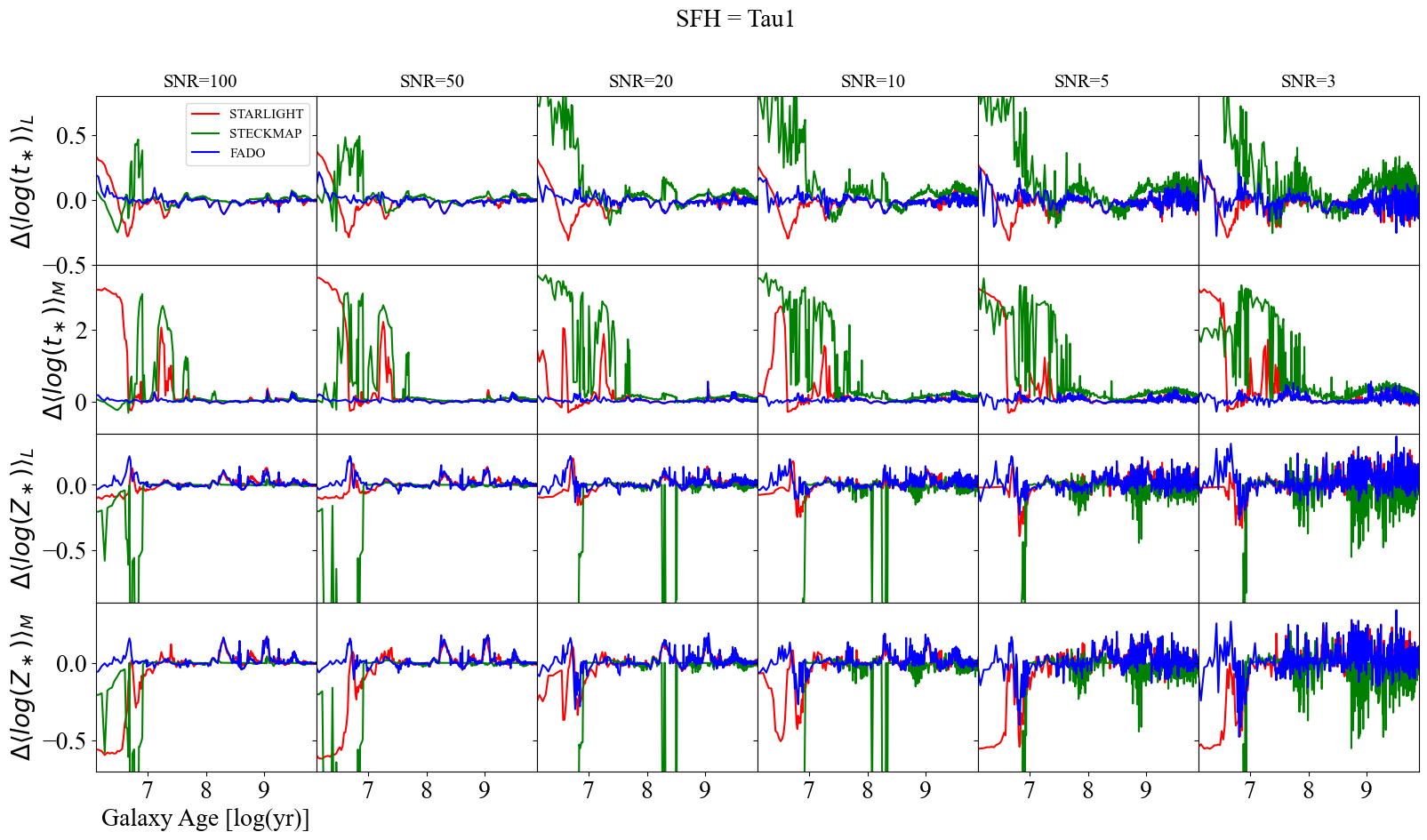}          
     \caption{Same caption of Fig. \ref{Cont-logtL} for \Tau\ SFH models. {\it Credits: Reproduced with permission from Astronomy \& Astrophysics, \textsuperscript{\textcopyright} ESO.}}\label{Tau1-logtL}
 \end{figure*}
 %--------------------------- 
 
\subsection{Instantaneous burst: \Tau\ models}

 Figure \ref{Tau1-logtL} reports the results obtained for the \Tau\ models, which show an overall consistency of the results, with $\Delta$\logtL\ $\sim$2-3\% for \fado\ and \starl\ already at S/N = 5. \stec\ results are dependent on the input S/N with unstable fits at $\log(t/\mathrm{yr}) > 7.5$.
 
 Mean metallicities are more homogeneous than \Con\ models, with unstable solutions at $\log(t/\mathrm{yr})\sim7$. In contrast to \fado\ and \starl, \stec\ fits show high instabilities at older ages above $\log(t/\mathrm{yr}) > 9$. Also for \Tau\ models mass-weighted quantities show a larger scatter. \fado\ shows $\Delta$\logtM\ of $\sim 0.1-0.2$ dex, while for \starl\ and \stec\ the uncertainties increase by a factor of $\sim$3.

\section{Conclusions}
\label{discussion}

For each tool considered the results accuracy depends not only on the input SFH but also on other factors, e.g. the S/N and the contribution of the nebular continuum. The accuracy of \logtL\ and \logtM\ changes drastically with the input SFH, with \Con\ models showing an increase in the uncertainties by a factor $\sim$3-5 from light- to mass-weighted quantities. Larger discrepancies are linked to phases where the nebular contribution is not negligible, which is particularly evident in \Con\ spectra. This confirms the importance of considering nebular emission for starburst systems, where this component can reach up to 50\% of the total optical and near-infrared emission \citep{kru,izo,pap,lei,sch}.
 
 Our work underlines the importance of using appropriate codes when considering the spectra from galaxies with high SFRs. A self-consistent treatment of nebular emission assumes a fundamental role for galaxies in starburst phases, which are more common at higher redshifts.
  
\acknowledgments This work was supported by Fundação para a Ci\^{e}ncia e a Tecnologia (FCT) through the research grants PTDC/FIS-AST/29245/2017, UID/FIS/04434/2019, UIDB/04434/2020 and UIDP/04434/2020.
%C. P. acknowledges support from DL 57/2016 (P2460) from the `Departamento de F\'{i}sica, Faculdade de Ci\^{e}ncias da Universidade de Lisboa'.
%L. S. M. C. acknowledges support by the project ``Enabling Green E-science for the SKA Research Infrastructure (ENGAGE SKA)" (reference POCI-01-0145-FEDER-022217), funded by COMPETE2020 and FCT. 
%P.P. acknowledges support by the project "Identifying the Earliest Supermassive Black Holes with ALMA (IdEaS with ALMA)" (PTDC/FIS-AST/29245/2017).

\bibliographystyle{aaabib}
%\bibliography{examplebib}
\bibliography{PappalardoBib}
\end{document}